\documentclass[dvipdfmx]{ws-mpla}
\usepackage[super]{cite}
\usepackage{graphicx}

\makeatletter
\def\@cite#1#2{$^{\hbox{\scriptsize{#1\if@tempswa , #2\fi}}}$}
\makeatother

\def\Journal#1#2#3#4{{#1} {\bf #2}, #3 (#4)}

\def\JCAP{J. Cosmol. Astropart. Phys.}

\def\JCAP{JCAP}

\def\MPLA{Mod. Phys. Lett. A}

\def\NPA{Nucl. Phys. A}
\def\NPB{Nucl. Phys. B}

\def\PDU{Phys. Dark. Univ.}
\def\PLA{Phys. Lett. A}
\def\PLB{Phys. Lett. B}
\def\PREP{Phys. Rep.}

\def\PLBOLD{Phys. Lett.}

\def\PRD{Phys. Rev. D}

\def\RPP{Rep. Prog. Phys.}

\begin{document}

\markboth{Jun Iizuka and Teruyuki Kitabayashi}
{Relativistic effective degrees of freedom and quantum statistics of neutrinos}

\catchline{}{}{}{}{}

\title{Relativistic effective degrees of freedom and quantum statistics of neutrinos}

\author{\footnotesize Jun Iizuka and Teruyuki Kitabayashi}

\address{Department of Physics, Tokai University, \\
4-1-1 Kitakaname, Hiratsuka, Kanagawa, 259-1292, Japan\\
teruyuki@keyaki.cc.u-tokai.ac.jp}

\maketitle

\pub{Received (Day Month Year)}{Revised (Day Month Year)}

\begin{abstract}
Analytical expressions of the relativistic effective degrees of freedom $g_\ast$ with non-pure fermionic neutrinos are presented. A semi-analytical study is performed to show that $g_\ast$ with pure fermionic neutrinos may be greater than $g_\ast$  with pure bosonic neutrinos for non-vanishing lepton flavor asymmetries.
\keywords{Neutrino statistics; Relativistic effective degrees of freedom}
\end{abstract}

\ccode{PACS Nos.14.60.St:98.80.Cq}

\section{Introduction}
\label{sec:intro}
We understand the neutrinos obey purely Fermi-Dirac statistics on the analogy of the electrons\cite{Bartalucci2006PLB}; however, they may possess mixed statistics \cite{Dolgov2005PLB,Barabash2007PLB,Choubey2006PLB,Ignatiev2006PLA,Tornow2010NPA,Vergados2012RPP,NEMO2014NPA}. Dolgov, et.al. studied the effects of continuous transition from Fermi-Dirac to Bose-Einstein statistics of neutrinos and discussed the possible modification of the big bang nucleosynthesis \cite{Dolgov2005JCAP}. J.I and T.K estimated the relativistic degrees of freedom with non-pure fermionic neutrinos in the early universe by numerical calculations \cite{Iizuka2015MPLA,Iizuka2016PDU}.  

In this letter, to complement our previous numerical studies \cite{Iizuka2015MPLA,Iizuka2016PDU}, we show analytical expressions of the relativistic effective degrees of freedom in the early universe with non-pure fermionic neutrinos, and perform a semi-analytical study by using these analytical expressions.

\section{Analytical expressions}
{\bf Net number density and net energy density: }
The distribution function is given by \cite{Dolgov2005JCAP}
\begin{eqnarray}
f_i=\frac{g_i}{e^{(E-\mu_i)/T}+\kappa_i},
\label{Eq:f_i}
\end{eqnarray}
where $g_i$, $E$, $\mu_i$ and $T$ denote the number of internal degrees of freedom, energy, chemical potential and temperature, respectively. The Fermi-Bose parameter $\kappa_i$ describes the continuous transition from Fermi-Dirac $\kappa_i=1$ distribution to Bose-Einstein $\kappa_i=-1$ distribution via Maxwell-Boltzmann $\kappa_i=0$ distribution. 

The net number density of particle species $i$ is obtained as \cite{Kolb1980NPB,Kolb1990}
\begin{eqnarray}
n_{i- \bar{i}}&=&n_i - n_{\bar{i}} \nonumber \\
&=&\frac{g_i}{2\pi^2}\int^\infty_{m_i}E(E^2-m^2_i)^{1/2} [f_i(E)-f_{\bar i}(E)]dE \nonumber \\
&\simeq& \frac{2g_i  T^3 \xi_i}{\pi^2}F_2(\kappa_i )
\label{Eq:n_i_ibar}
\end{eqnarray}
where 
\begin{eqnarray}
\xi_i=\frac{\mu_i}{T}, \quad F_s(x) \equiv \frac{{\rm Li}_s(-x)}{-x},
\end{eqnarray}
and ${\rm Li}_s (x)$ denotes the polylogarithm function. We note that $F_2(1)=\pi^2/12$, $F_4(1)=7\pi^4/720$ are obtained for pure fermions and $F_2(-1)=\pi^2/6$, $F_4(-1)=\pi^4/90$ are obtained for pure bosons. 

Similarly, the net energy density is obtained as
\begin{eqnarray}
\rho_{i + \bar{i}} = \rho_i + \rho_{\bar{i}} \simeq \frac{3g_i T^4}{\pi^2}\left(2F_4(\kappa_i)+\xi_i^2F_2(\kappa_i)\right).
\label{Eq:rho_i_bar_i}
\end{eqnarray}
%

{\bf Chemical potentials: }
The beta-decay of down quark via weak interactions $d \rightarrow u + \ell + \bar{\nu}_\ell$ provides 
\begin{eqnarray}
\mu_u+\mu_\ell&=&\mu_d+\mu_{\nu_\ell}, \quad (\ell=e,\mu,\tau),
\label{Eq:beta_decay}
\end{eqnarray}
and the relation of $\mu_i =\mu_{\bar i}$ and $\mu_\gamma=\mu_W=\mu_Z=\mu_g=\mu_H=0$ are appropriate. With the following assumptions \cite{Stuke2012JCAP}
\begin{eqnarray}
\mu_u =\mu_c =\mu_t, \quad
\mu_d =\mu_s=\mu_b, 
\end{eqnarray}
there are only five independent chemical potentials and we take these as $\mu_u, \mu_d,\mu_{\nu_e}, \mu_{\nu_\mu}, \mu_{\nu_\tau}$. These five independent chemical potentials are uniquely determined by the following five conservation laws \cite{Stuke2012JCAP}
\begin{eqnarray}
    sQ&=&-\sum_{i=e,\mu,\tau}n_{i-\bar{i}} + \frac{2}{3}\sum_{i=u,c,t}n_{i-\bar{i}}-\frac{1}{3}\sum_{i=d,s,b}n_{i-\bar{i}}, \label{Eq:sq} \\
sB&=&\frac{1}{3}\sum_{i=quarks}n_{i-\bar{i}}, \label{Eq:sb} \\
sL_\ell &=& n_{\ell-{\bar{\ell}}} +n_{\nu_\ell-{\nu}_{\bar{\ell}}}, \label{Eq:sell} 
\end{eqnarray}
where $Q$, $B$ and $L_\ell$ denote electric charge, baryon number and lepton flavor number of the universe, respectively.

For electrically neutral universe, $Q=0$, from Eqs. (\ref{Eq:n_i_ibar}), (\ref{Eq:sq}), (\ref{Eq:sb}) and (\ref{Eq:sell}), the following coupled equations for chemical potentials up to $\mathcal{O}(\xi_i^2)$ are obtained
\begin{eqnarray}
&&0 = -(\xi_e+\xi_{\mu}+\xi_{\tau}) +4\xi_{u}-3\xi_{d}, \nonumber \\
&&\frac{3sB}{T^3} = 2\xi_{u}+3\xi_{d},\nonumber  \\
&&\frac{3sL_\ell}{T^3} = \xi_\ell+\frac{6}{\pi^2}F_2(\kappa_\nu)\xi_{\nu_\ell},
\label{Eq:from_Q_0}
\end{eqnarray}
where we assume $m_W < T < m_t$, $\kappa_\nu=\kappa_{\nu_e}=\kappa_{\nu_\mu}=\kappa_{\nu_\tau}$ and take $\kappa_i=1$ for the fermions in the standard model except neutrinos. From Eqs.(\ref{Eq:beta_decay}) and (\ref{Eq:from_Q_0}), the five independent  $\xi_i$ ($i=u, d,\nu_e,\nu_\mu,\nu_\tau$) as well as $\xi_\ell$ are analytically determined
\begin{eqnarray}
\label{Eq:xi_u_d_nu_ell}
\xi_u&=&\frac{s}{2T^3}\frac{1}{1+\frac{11}{\pi^2}F_2(\kappa_\nu)}\left\{L+\left[1+\frac{12}{\pi^2}F_2(\kappa_\nu)\right]B\right\}, \\
\xi_d&=&\frac{s}{3T^3}\frac{1}{1+\frac{11}{\pi^2}F_2(\kappa_\nu)}\left\{-L+\left[2+\frac{21}{\pi^2}F_2(\kappa_\nu)\right]B\right\},\nonumber \\
\xi_{\nu_\ell}&=&\frac{s}{T^3}\frac{1}{1+\frac{6}{\pi^2}F_2(\kappa_\nu)}\left\{3L_\ell+\frac{1}{6(1+\frac{11}{\pi^2}F_2(\kappa_\nu))}\left[5L-\left(1+\frac{6}{\pi^2}F_2(\kappa_\nu)\right)B\right]\right\},
\nonumber \\
\xi_{\ell}&=&\frac{s}{T^3}\frac{1}{1+\frac{6}{\pi^2}F_2(\kappa_\nu)}\left\{3L_\ell-\frac{F_2(\kappa_\nu)}{\pi^2(1+\frac{11}{\pi^2}F_2(\kappa_\nu))}\left[5L-\left(1+\frac{6}{\pi^2}F_2(\kappa_\nu)\right)B\right]\right\},\nonumber 
\end{eqnarray}
where $L=\sum_{\ell=e,\mu,\tau} L_\ell$.  

{\bf Relativistic effective degrees of freedom:}
The relativistic effective degrees of freedom for energy density, $g_*$, is defined by \cite{Kolb1990,Stuke2012JCAP}
\begin{eqnarray}
\rho=\sum_i \rho_i =\frac{\pi^2T^4}{30} g_\ast.
\label{Eq:rho_gast}
\end{eqnarray}
Similarly, for entropy density, $g_{\ast s}$ is defined by $s=\sum_i s_i =\frac{2\pi^2T^3}{45} g_{\ast s}$. From Eqs.(\ref{Eq:rho_gast}), (\ref{Eq:rho_i_bar_i}) and (\ref{Eq:xi_u_d_nu_ell}), the relativistic effective degrees of freedom for energy density is obtained as
\begin{eqnarray}
g_*(\xi,\kappa_\nu)=  g_*(\xi_i=0,\kappa_\nu=1) +\Delta g_*,
\end{eqnarray}
where 
\begin{eqnarray}
g_*(\xi_i=0,\kappa_\nu=1)  = \sum_{b=bosons}g_b + \frac{7}{8} \sum_{f=fermions} g_f,
\end{eqnarray}
denotes the well-known relativistic effective degrees of freedom for vanishing chemical potentials ($\xi_i=0$) and for pure fermionic neutrinos ($\kappa_\nu=1$) \cite{Kolb1990}. The effects on $g_\ast$ from the non-vanishing chemical potentials and non-pure fermionic neutrinos are estimated as 
\begin{eqnarray}
\Delta g_*&=& \frac{15}{4\pi^2} \sum_{i=fermions\neq \nu} g_i \xi_i^2(\kappa_\nu)+ \sum_{i=\nu}\frac{45}{\pi^4}F_2(\kappa_\nu)g_\nu \xi_i^2 (\kappa_\nu)\nonumber \\
&&+\sum_{i=\nu}g_i\left( \frac{90}{\pi^4} F_4(\kappa_\nu) - \frac{7}{8} \right).
\label{Eq:Delta_g}
\end{eqnarray}

Similarly, the relativistic effective degrees of freedom for entropy density is expressed as $g_{*s}(\xi,\kappa_\nu)=  g_{*s}(\xi_i=0,\kappa_\nu=1) +\Delta g_{*s}$. For $T \sim 100$GeV, we obtain $g_{*s}(\xi_i=0,\kappa_\nu=1)=g_{*}(\xi_i=0,\kappa_\nu=1)$ and $\Delta g_{*s}=\Delta g_{*}$.

\section{Discussions and summary}
With the vanishing chemical potential, the equilibrium energy density of pure bosonic particle is larger than it of pure fermionic particle \cite{Kolb1990}. One may expect that the relation $g_*^{\rm FD}<g_*^{\rm BE}$ is guaranteed where $g_\ast^{\rm FD}$ denotes $g_\ast$ with pure fermionic neutrinos and $g_\ast^{\rm BE}$ denotes $g_\ast$ with pure bosonic neutrinos. However, in our previous numerical studies\cite{Iizuka2015MPLA,Iizuka2016PDU}, we have shown that this relation is not always satisfied with non-vanishing lepton flavor asymmetries in the early universe \cite{Steigman1977PLB,Dolgov2002NPB,Abazajian2002PRD,Wong2002PRD,Mangano2005NPB,Schwarz2009JCAP,Mangano2012PLB,Ichimasa2014PRD}. 

We complement our previous numerical studies \cite{Iizuka2015MPLA,Iizuka2016PDU} by semi-analytical calculations. From Eqs. (\ref{Eq:xi_u_d_nu_ell}) and (\ref{Eq:Delta_g}), we obtain   
\begin{eqnarray}
\Delta g_*(\kappa_\nu=1)=\frac{50}{529\pi^2}\left( \frac{s}{T^3}\right)^2 \left(873B^2 -162BL+362L^2+1587 \sum_\ell L_\ell^2 \right),
\end{eqnarray}
for pure fermionic neutrinos, and
\begin{eqnarray}
\Delta g_*(\kappa_\nu=-1)=\frac{45}{289\pi^2}\left( \frac{s}{T^3}\right)^2 \left(529B^2 -78BL+140L^2+867 \sum_\ell L_\ell^2 \right)
+\frac{3}{4},
\end{eqnarray}
for pure bosonic neutrinos. For the sake of simplicity, we assume $L_\ell=L/3$ and $L \gg B \sim 0$, and use very rough estimation of $s/T^3 = 2\pi^2 g_{*s}/45=2\pi^2 g_{*}/45 \sim 44$ with $T=100$ GeV. In this case, we obtain
\begin{eqnarray}
\Delta g_*(\kappa_\nu=1) \simeq 1.65 \times 10^{4} L^2, \quad 
\Delta g_*(\kappa_\nu=-1) \simeq 1.31 \times 10^{4} L^2+\frac{3}{4}.
\end{eqnarray}
and $\Delta g_*(\kappa_\nu=1) \gtrsim \Delta g_*(\kappa_\nu=-1)$ as well as $g_*^{\rm FD} \gtrsim g_*^{\rm BE}$ for $L \gtrsim 0.015$. 

We comment on possible application of our results in a cosmological context. In the leptogenesis scenario \cite{Davidson2008PREP}, the baryon-photon ratio in the universe $\eta_B$ is related to the lepton asymmetry $Y_L$ via $\eta_B \propto Y_L \propto g_\ast^{-1}$. Thus, the lepton number in the early universe yields change of the baryon-photon ratio. More detailed analysis will be found in our future study.

In summary, analytical expressions of the relativistic effective degrees of freedom with non-pure fermionic neutrinos are presented. A semi-analytical study has been performed to complement our previous numerical studies which show that the relation of $g_*^{\rm FD} \gtrsim g_*^{\rm BE}$  may be allowed with non-vanishing lepton flavor asymmetries.



\end{document}